\documentstyle[12pt]{article}
\newfont{\bigfont}{cmr17 scaled 3583}
\newfont{\bfont}{cmr17 scaled 2000}
\def\Bigs{\mathop{\hbox{\bigfont\symbol{'006}}}\limits}
\def\bigs{\mathop{\hbox{\bfont\symbol{'006}}}\limits}
\def\bprod{\mathop{\hbox{\bfont\symbol{'005}}}\limits}
\def\Bigsum#1#2{\displaystyle \Bigs_{#1}^{#2}}
\def\bigsum#1#2{\raisebox{-1ex}{$\displaystyle \bigs_{#1}^{#2}$}}
\def\bigprod#1#2{\raisebox{-1ex}{$\displaystyle \bprod_{#1}^{#2}$}}
\def\pn{\par\noindent}
\def\bn{\par\bigskip\noindent}
\def\del{\partial}
\def\vpo{\varphi_1}
\def\vpt{\varphi_2}
\def\abs#1{\mid\! #1 \!\mid}
\def\BBCT{\,\hbox{\hbox to -2.8pt{\vrule height 6.7pt width .3pt
  \hss}\rm C}}
\def\BBCS{\,\hbox{\hbox to -2.2pt{\vrule height 4.5pt width.2pt
  \hss}$\scriptstyle\rm C$}}
\def\BBCSS{\,\hbox{\hbox to -2pt{\vrule height 3.3pt width .2pt
  \hss}$\scriptscriptstyle \rm C$}}
\def\BBC{{\mathchoice{\BBCT}{\BBCT}{\BBCS}{\BBCSS}}}
\def\BZT{{\rm Z{\hbox to 3pt{\hss\rm Z}}}}
\def\BZS{{\hbox{$\scriptstyle\rm Z${\hbox to 1.8pt
         {$\hss\scriptstyle\rm Z$}}}}}
\def\BZSS{{\hbox{$\scriptscriptstyle\rm Z${\hbox to 1.8pt
          {$\hss\scriptscriptstyle\rm Z$}}}}}
\def\BZ{{\mathchoice{\BZT}{\BZT}{\BZS}{\BZSS}}}
\renewcommand{\theequation}{\mbox{\arabic{section}.\arabic{equation}}}
\newcommand{\reseteqn}{\setcounter{equation}{0}}
\newcommand{\mysection}{\reseteqn\section}

\begin{document}
\thispagestyle{empty}
\begin{raggedleft}
BONN-Th-94-22\\
hep-th/yymmxxx\\
Nov.\ 1994\\
\end{raggedleft}
$\phantom{x}$\vskip 0.618cm\par
{\huge \begin{center}
Topology and Fractional Quantum Hall Effect
\end{center}}\par
\vfill
\begin{center}
$\phantom{X}$\\
{\Large Raimund Varnhagen\footnote[1]{Supported by the Deutsche
Forschungsgemeinschaft}}\\[3ex]
{\em Physikalisches Institut\\ der Universit\"at Bonn\\
Nussallee 12\\
D-53115 Bonn\\ Germany\\
email: varnhagen@uni-bonn.de
}
\end{center}\par
\vfill
\begin{abstract}
  \noindent
  Starting from
  Laughlin type wave functions with generalized periodic boundary
  conditions describing the degenerate groundstate of a quantum Hall
  system we explictly construct $r$ dimensional vector bundles. It turns
  out that the filling factor $\nu$ is given
  by the topological quantity $c_1 \over
  r$ where $c_1$ is the first chern number of these vector bundles.
  In addition, we managed to proof that under physical natural
  assumptions the stable vector bundles correspond to the
  experimentally dominating series of measured fractional
  filling factors
  $\nu = {n \over 2pn\pm 1}$. Most remarkably, due to the very special
  form of the Laughlin wave functions the fluctuations of the curvature
  of these vector bundles converge to zero in the limit of infinitely
  many particles which shows a new mathematical property. Physically,
  this means that in this limit the Hall
  conductivity is independent of the boundary conditions which is very
  important for the observabilty of the effect. Finally we discuss the
  relation of this result to a theorem of Donaldson.
\end{abstract}
\vfill
%email: varnhagen@uni-bonn.de
\newpage
\setcounter{page}{1}
\mysection{Introduction}
\pn
Since its discovery the integer and fractional Quantum Hall effect (IQHE,
FQHE) \cite{KDP80,TSG82}
have fascinated both experimentalists and theorists.
The conductance of a two
dimensional electron gas in a high magnetic field at low temperatures
exhibits quantized plateau values of the form
$\sigma_{xy} = {e^2 \over h} \nu$ where
the filling fraction $\nu$ is an integer or fractional number. It was very
astonishing that the conductance  does not depend on the size and
geometry of the experimental sample and that the quantization could be
measured with such a high precision.
In the last few years, it became more and more obvious that in
completely spin-polarized quantum Hall systems
the fractional values
\begin{eqnarray}
  \nu={n \over 2pn \pm 1}\qquad {\rm and}\qquad
  \nu=1-{n \over 2pn \pm 1}
  \label{ms}
\end{eqnarray}
first proposed by J.K.~Jain \cite{J89a} turn out to
dominate the experimental measurements of the FQHE for $\nu<1$
\cite{DST93,WRP93,HLR93}.
\footnote{In addition to these fractional values
other fractional plateaus are measured but without vanishing
of the longitudinal conductivity. This has been explained by
J.~Fr\"ohlich et.~al.~ \cite{FST94} by a classification of quantum
Hall fluids}. \pn
{}From a theoretical point of view there are two main questions
that have to be solved in order to understand
the QHE: a) What is the reason for the exact
quantization of the Hall conductance especially, for the series
(\ref{ms})?
b) Why do Hall plateaus occur? The aim of this letter is to give some
insight to the first problem, especially to understand the appearance
of the experimentally observed main series (\ref{ms}) by means of
topological arguments. The second question is not considered here,
but it is widely suspected that disorder effects are responsible for the
occurrence and the width of the Hall plateaus, which are essential for
observing the IQHE as well as the FQHE \cite{PG87}. \pn
Historically, first arguments
for the exact quantization of the Hall conductance were given
by Laughlin who showed that the Hall conductance is
quantized whenever the Fermi energy lies in an energy gap,
even if the gap lies within a Landau level \cite{L81}.
A very
important contribution to an understanding of the IQHE was provided
by Thouless et.\ al., deriving the Hall conductance from the Kubo formula
via linear response theory and proving the locality  of the Hall conductance
inside the sample; thus, the main consequence of this work is
that the Hall conductance is essentially
insensitive to the boundary conditions one imposes at the edges of the
physical samples \cite{TKN82,NT87}.
J.E.~Avron et.\ al. and M.~Kohmoto performed the topological
analysis of the IQHE
in a convincing way and showed that the Hall conductance
can be represented by the first Chern number of a line bundle over the
magnetic Brillouin zone which is an integer-valued topological constant
\cite{ASS83,AS85,K85,N81}.
All these considerations assume that the groundstate is nondegenerate.
\pn
In the case of the FQHE there are many completely different theoretical
descriptions. The aim of that paper is to show the connection of two
of them. \pn
On the one hand, Thouless et.\ al.\ required a degeneracy in order to
explain the fractional quantization,
since a nondegenerate groundstate always leads to an integral
quantization.
Consequently the Hall conductance has to be expressed
by the first Chern number of a vector bundle divided by its rank, where
the rank is equal to the degeneracy \cite{NTW85}. \pn
On the other hand, an important step was taken by Laughlin who wrote
down
the wave functions for the fundamental fractions $\nu = {1\over 3},
{1\over 5},{1\over 7},\ldots$ \cite{L83}.
Extensive calculations have proven these wave functions to be extremely
close to the numerical exact solutions \cite{PG87}.
They play a special role in a hierarchal scheme in
which a daughterstate is obtained at each step from a condensation of
quasiparticles of the parent state into a correlated low energy state
\cite{H83,H84}.
\pn
Laughlin's wave functions defined on a disk or sphere do
not show any degeneracy of the groundstate.
However, this degeneracy naturally
occurs if the wave functions are defined on a torus, that means if
generalized periodic boundary conditions are imposed.
First F.D.M.~Haldane
and E.H.~Rezayi \cite{HR85} and later E.~Keski-Vakkurri and X.G.~Wen
\cite{KW93}
have constructed Laughlin type wave functions with periodic boundary
conditions which we slightly generalized.\pn
This paper is organized as follows: In the next two sections we briefly
review the two theoretical descriptions of the FQHE. Firstly we show how
to derive the topological quantities from the Kubo formula, secondly we
recall Laughlin type wave functions with
and without periodic boundary conditions and generalize the known results.
In the fourth section we show how both concepts fit
together, constructing vector bundles from these wave functions. It turns
out that the experimentally measured main series (\ref{ms}) is described
by stable vector bundles. Further, we calculate the fluctuations of the
Hall conductance which vanish in the limit of infinite number of
particles due of the special form of the Laughlin wave functions. From
a physical point of view this explains the independence of the Hall
conductivity from the boundary conditions; from a mathematical point of
view this is a necessary condition for stability of a vector bundle due
to a theorem of S.K.~Donaldson \cite{D83}.
\bn
\bn
\mysection{From the Kubo formula to Chern Numbers}
Following Thouless et.\ al.,
we consider a two dimensional interacting electron system in both a
magnetic field perpendicular to the plane of area $A$
and an electric field in the $x$-direction of the plane.
Such a system is described by an N-body Hamiltonian
\begin{eqnarray}
  H = \sum_{i=1}^N {1 \over 2m}\left( \vec p_i-{e\over c}\vec A
  \right)^2 + \sum_{i=1}^N U(x_i,y_i) + \sum_{i<j}
   V(\mid r_i-r_j\mid)\,.
  \label{Hamiltoian}
\end{eqnarray}
The Hall current which flows in the $y$-direction
can be obtained via linear response theory by slowly switching on
an electric field in $x$-direction. The first order perturbation
expresses the Hall conductivity by the Kubo formula:
\begin{eqnarray}
  \sigma_{xy}={e^2\hbar \over i A}\sum_{E^{\alpha}<E_F<E^{\beta}}
  {(v^y)_{\alpha\beta}(v^x)_{\beta\alpha}-
   (v^x)_{\alpha\beta}(v^y)_{\beta\alpha} \over(E^{\alpha}-E^{\beta})^2}
  \label{Kubo}
\end{eqnarray}
where $E_F$ is a Fermi energy and the summation implies the sum over
all states below and above the Fermi energy. The indices $\alpha$
and $\beta$ label the bands of the N-body Hamiltonian in the absence of
the external electric field.
The velocity operator appearing in the Kubo formula is given by
\begin{eqnarray}
   \vec v = \sum_{i=1}^N {1 \over m} (\vec p_i -{e\over c}\vec A)\, .
\end{eqnarray}
The Hamiltonian has a symmetry of magnetic translations which are
generated by
\begin{eqnarray}
   k^x_i &=& p^x_i-{e\over c}A^x - {e\over c}By_i \\
   k^y_i &=& p^y_i-{e\over c}A^y + {e\over c}Bx_i\qquad {\rm with}\\
   \left[ k_i^x,k_i^y \right]  &=& \frac{\hbar e}{i c}B \,,
\end{eqnarray}
where the magnetic field is given as $B=\del_x A_y-\del_y A_x$.
Then, the magnetic translations are defined by
\begin{eqnarray}
   t_i(\vec L) &=& \exp({i \over \hbar}{\vec k_i}\cdot{\vec L})
   \qquad {\rm with} \\
   t_i(\vec a)t_i(\vec b) &=& t_i(\vec b)t_i(\vec a)
   e^{-i(a\times b)/l^2}
\end{eqnarray}
where $l^2={\hbar c\over eB}$ is the fundamental length of the system.
Thus, the many-body magnetic translation
\begin{eqnarray}
   T(\vec a) \equiv \prod_{i=1}^{N} t_i(\vec a)
\end{eqnarray}
commute with the Hamiltonian for appropriate potential $U(x,y)$.\pn
In order to utilize this symmetry, we impose on the many-body wave function
the generalized periodic boundary conditions
\begin{eqnarray}
   t_i(L_1)\psi &=& e^{2\pi i\vpo}\psi \\
   t_i(L_2)\psi &=& e^{2\pi i\vpt} \psi\,,
   \label{pb}
\end{eqnarray}
where the parameters $\vpo$ and $\vpt$ are independent of
the particles
indices, as required by the total antisymmetry of the wave function.
Now, we perform the gauge transformation
\begin{eqnarray}
   \psi &\longrightarrow& \exp\Big(-{i\over \hbar}
   \tilde\vpo(x_1+\ldots + x_N)\Big)\\
   &&\quad\times\exp\Big(-{i\over \hbar}
   \tilde\vpt(y_1 +\ldots + y_N)\Big) \psi \\
   p^x_i &\longrightarrow & p^x_i + \tilde\vpo \\
   p^y_i &\longrightarrow & p^y_i + \tilde\vpt \,.
\end{eqnarray}
It is clear that ${1\over \hbar}\del\tilde H/\del\tilde\vpo$
and ${1\over\hbar}\del\tilde H/\del\tilde\vpt$
are just the transformed velocity operators of the Kubo formula
(\ref{Kubo}) where $\tilde H$ is the transformed Hamiltonian $H$.
Thus, after some manipulations the Kubo formula can be written as
\begin{eqnarray}
  \sigma_{xy}={e^2\over i\hbar}\sum_{E_{\alpha}<E_F<E_{\beta}}
  \left( \langle {\del\psi^{\alpha}\over\del\vpt}\mid\beta\rangle
  \langle\beta\mid {\del\psi^{\alpha}\over\del\vpt} \rangle -
  \langle {\del\psi^{\alpha}\over\del\vpo}\mid\beta\rangle
  \langle\beta\mid{\del\psi^{\alpha}\over\del\vpo} \rangle \right)\,.
\end{eqnarray}
Under the assumption that the groundstate $\psi_0$
is nondegenerate and that regardless of the boundary
conditions, there exists a finite energy gap between the groundstate
and the exited bulk states -- this agrees with the experimentally
observed fact that the longitudinal conductance $\sigma_{xx}$ vanishes
at each Hall plateau, since $\sigma_{xx}$ should be proportional to
$\exp\left( {-\Delta\over kT}\right)$ where $\Delta$ is the value of the
energy gap -- the conductance can be further simplified:
\begin{eqnarray}
  \sigma_{xy}={e^2 \over \hbar i}\left(\langle {\del \psi^0\over \del
  \vpo}\mid{\del \psi^0\over \del \vpt}\rangle - \langle {\del
  \psi^0\over \del \vpt} \mid {\del \psi^0\over \del \vpo}\rangle
  \right)\,.
\end{eqnarray}
So far, we have just performed formal transformations of the Hall
conductance without considering its quantization. To understand
the latter, we have to equate $\sigma_{xy}$ with its average
over all the phases that specify different boundary conditions.
\begin{eqnarray}
  \sigma_{xy}={e^2 \over h} {1 \over 2\pi i}\int_{0}^{2\pi} \int_{0}^{2\pi}
  d\vpo d\vpt \left(\langle {\del \psi^0\over \del
  \vpo}\mid{\del \psi^0\over \del \vpt}\rangle - \langle {\del
  \psi^0\over \del \vpt} \mid {\del \psi^0\over \del \vpo}\rangle
  \right) \,.
  \label{line bundle}
\end{eqnarray}
This integral is actually a topological invariant. It is the first Chern
number
of a $U(1)$ line bundle of the groundstate wave function on the base
manifold of a Torus $T^2$ parameterized by $\vpo$ and $\vpt$.
Thus, one can draw the important conclusion: on condition that the
groundstate is nondegenerate and there exists a finite energy gap, the
Hall conductance is always quantized. \cite{TKN82,ASS83,AS85,K85,NT87}
\pn
If one generalize to a degenerate groundstate of degeneracy $r$,
the equation $(\ref{line bundle})$ has to be replaced by:
\begin{eqnarray}
  \sigma_{xy}={e^2 \over h} {1\over r}\sum_{i=1}^{r} {1 \over 2\pi i}
  \int_{0}^{2\pi} \int_{0}^{2\pi}
  d\vpo d\vpt \left(\langle {\del \psi_i^0\over \del
  \vpo}\mid{\del \psi_i^0\over \del \vpt}\rangle - \langle {\del
  \psi_i^0\over \del \vpt} \mid {\del \psi_i^0\over \del \vpo}\rangle
  \right) \,,
  \label{vector bundle}
\end{eqnarray}
where $\{ \psi_i\}$ is an orthogonal basis spanning the groundstate Hilbert
space \cite{NTW85}.
This integral again is a topological invariant, namely, the first
Chern number of the determinant bundle of the vector bundle
of rank $r$ which is given by this groundstate
Hilbert space on the same base manifold as in the nondegenerate case.
Thus, the Hall conductance is given as a fraction
\begin{eqnarray}
  \sigma_{xy} = {e^2 \over h} {c_1 \over r} \,,
\end{eqnarray}
where $c_1$ is the first Chern number of the above vector bundle.
The disadvantage
of this approach to the FQHE is that neither the Laughlin
wave function for a disk, nor Haldane's equivalent expression for the
wave function on a spherical surface appear to have any degeneracy.
Furthermore, these degeneracies
are not independent of impurities and disorder. The independence should
be respected because impurities an disorder
are generally believed to be as essential to the observability of the
fractional effect as it is to the integer effect. \pn
%%%%%%%%%%%%%%%%%%%%%%%%%%%%%%%%%%%%%%%%%%%%%%%%%%%%%%%%%%%%%%%%%%%%%%%%%%
\mysection{Laughlin wave functions}
In this section we consider the problem of electrons moving in a
two dimensional surface in the presence of a perpendicular magnetic field
from another point of view following Laughlin \cite{L83,PG87}.
Let us first consider the single electron
Hamiltonian
\begin{eqnarray}
  H = {1 \over 2m}(\vec p - {e\over c} \vec A)^2 = {1\over 2}m
  (v_x^2+v_y^2)
\end{eqnarray}
with constant magnetic field $B=\del_xA_y-\del_yA_x$.
Since the velocity operator
\begin{eqnarray}
  \vec v = {1\over m}(\vec p -{e\over c}\vec A)
\end{eqnarray}
fulfills the following commutation relation
\begin{eqnarray}
    [v_x,v_y]={\hbar \omega_c \over m}i \qquad{\rm with}\qquad
    \omega_c={e B\over m c}
\end{eqnarray}
the energy spectrum of this Hamiltonian is the same as that of the
harmonic oscillator:
\begin{equation}
   E_n=\hbar\omega_c(n+{1\over 2})\,.
\end{equation}
Each level, called Landau level, is infinitely degenerate due to
the rotational symmetry around the $z$-axes. In a finite sample of area
$A$, one can show that the degeneracy of each Landau level is determined
by the number of magnetic flux quanta
\begin{equation}
    N_S = {B A\over \Phi_0} = {\Phi_{mag} \over \Phi_0}
\end{equation}
where $\Phi_{mag}$ is the magnetic flux through the area $A$ and $\Phi_0
={h c\over e}$ is a single flux quantum.
In the Landau gauge the wave functions in the lowest Landau level (LLL)
$(n=0)$ are generated by
\begin{equation}
   \psi(z) \sim z^k \exp({1\over 4l^2}\abs{z}^2)\,,\qquad
    k=0\ldots N_S-1\,,
   \qquad z=x+iy\qquad l^2={\hbar c\over e B}\,.
   \label{LLL}
\end{equation}
Let us now consider the case of $N$ such electrons. If there is no
interaction  between them, the many-particle problem splits into $N$
copies of the single particle problem. Since the magnetic field $B$
controls the number of states and thus the density of electrons per
state, its action can be considered as an external pressure. The
electron density per state i~.e.~ the filling factor is defined as
$\nu$:
\begin{equation}
   \nu={N\over N_S}\,.
\end{equation}
Laughlin was the first to realize that the many-particle groundstate of
the QHE is fundamentally different from other kown condensed states,
describing
magnetism or superconductivity. For the FQHE, with filling fraction
$\nu={1\over 2p+1}$ he found by numerical experiments, the groundstates
to be given by the following wave functions:
\begin{eqnarray}
   \psi(z_1,\ldots,z_N) \sim  \prod_{i<j}(z_i - z_j)^{2p+1}
            \exp(-{1\over 4l^2}\sum_i \abs{z_i}^2)
  \label{LW}
\end{eqnarray}
where $p$ should be an integer in order to make $\psi$
obey the Pauli principle.
The filling factor $\nu$ for these wave functions can be determined with
the help of the following argument from (\ref{LLL}):
the highest power of $z_i$ plus one is exactly the number of possible
states which can be occupied by one electron. In the case of the
Laughlin wave functions (\ref{LW}) this is
$N_S=(2p+1)(N-1)+1$ for each electron $z_i$. In the limit of
infinitely many
particles the filling fraction $\nu$ converges to $1/(2p+1)$.
Extensive calculations have proven these wave functions to be extremely
close to the numerical exact solution.\pn
Generalizations of  these wave functions to other filling fractions
exist \cite{WZ92,J89b,S91,W92}.
These are Laugh\-lin type wave functions
of the following form:
\begin{eqnarray}
    \psi_K(\{z_i^I\}) \sim \prod_{I \atop i<j }(z_i^I-z_j^I)^{K_{II}}
             \prod_{I<J\atop i\leq j} (z_i^I - z_j^J)^{K_{IJ}}
              \exp(-{1 \over 4l^2} \sum_{I,i} \abs{z_i^I}^2)
    \label{wf}
\end{eqnarray}
where now the electrons are distributed to $n$ different subbands which
are labeled by $I,J=1\ldots n$ each subband $I$ containing $N_I$
electrons.
To assume the validity of the Pauli principle and the single valuedness
of the wave function
$K_{IJ}$ should be a symmetric, positive, integer valued matrix with odd
integers on the main diagonal. These different subbands
are interpreted as different layers or different Landau levels or as
additional quantum numbers in the first Landau level, which can also
depend on the third coordinate.
\pn
The highest power of $z_i^I$ plus one gives the number of states which
can be occupied by the electron:
\begin{eqnarray}
    K_{II}(N_I - 1) + \sum_{I\neq J} K_{IJ}N_J = \sum_J K_{IJ}N_J - K_{II}
\end{eqnarray}
The magnetic field determines the number of possible states, which
is exactly the number of flux quanta $N_{\Phi}$, so:
\begin{eqnarray}
  N_{\Phi} = \sum_J K_{IJ}N_J - K_{II} \longrightarrow \sum_J K_{IJ}N_J
  \qquad {\rm for\ } N_I,N_{\Phi} {\rm\ large}
  \label{Kcond}
\end{eqnarray}
It follows: $ N_I = \sum_J (K^{-1})_{IJ} N_{\Phi} $.
Thus, the filling fraction is now given by:
\begin{eqnarray}
   \nu = { \sum_I N_I \over N_{\Phi}} = \sum_{IJ} (K^{-1})_{IJ}
   \label{ff}
\end{eqnarray}
F.D.~Haldane and E.H.~Rezayi have constructed Laughlin's wave functions
in the Landau gauge with periodic boundary conditions
for the filling fraction $\nu = {1\over m}$. Later E.~Keski-Vakkuri
and X.G.~Wen have generalized this to the Laughlin type wave functions
of the form (\ref{wf}). \pn
Following Haldane and Rezayi,
the wave function describing a particle confined to the lowest Landau
level has the analytic form in the Landau gauge ($\vec A=-Bye_x$)
\begin{equation}
   \psi(x,y)=\exp(-{1\over 2l^2}y^2)f(z),\qquad z=x+iy
\end{equation}
where $f(z)$ is a holomorphic function. The periodic boundary condition
on the wave function (\ref{pb}) is given by
(with $\tau=L_2e^{i\theta}$)
\begin{eqnarray}
   f(z+L_1)           &=&  e^{2\pi i\vpo}f(z)\\
   f(z+L_2e^{i\theta})&=&  e^{2\pi i\vpt}\exp(i\pi N_s[(2z/L_1)+\tau])f(z)
\end{eqnarray}
Since $f(z)$ is a holomorphic function, the integral of ${f'(z)\over f(z)}$
along the boundaries requires the number of zeros to be precisely $N_s$.
The possible analytic form of $f(z)$ is thus strongly constrained, and the
most general form is expressible as:
\begin{eqnarray}
   f(z)&=&\exp(ikz)\prod_{n=1}^{N_s}
   \vartheta_1\left((z-z_n)/L_1\mid \tau\right)\\
   \vartheta_1(z\mid\tau)&=& \sum_{n}\exp(i\tau\pi(n+{1\over2})+2\pi i
   (n+{1\over 2})(z+{1\over 2}))\,.
\end{eqnarray}
There are some additional constraints on $k$ and the $z_n$
that guarantee the number of linearly independent solutions to be
equal to the number of zeros of $f(z)$ \cite{HR85}.\pn
Let us now consider the many-particle wave functions. F.D.M.~Haldane and
E.H.~Rezayi have first used translational invariance to express those as a
product of a center-of-mass term and a factor involving only relative
coordinates. Following R.B.~Laughlin they have written down the
groundstate wave function for $\nu=1/m$ with periodic boundary
conditions (\ref{pb}); this has been generalized to other rational
filling factors by E.~Keski-Vakkuri and X.G.~Wen for periodic boundary
conditions with $\vpo=0$ and $\vpt=0$:
They expressed the groundstate wave function in the following way:
\begin{eqnarray}
 \psi_K(\{z_i^I\})\!\!&\!\!=\!\!&\!\!F^{c.m.}(Z_1,\ldots,Z_N)
     \prod_{I\atop i<j }\vartheta_1(z_i^I-z_j^I)^{K_{II}}
     \prod_{I<J\atop i\leq j}\vartheta_1(z_i^I - z_j^J)^{K_{IJ}}\\
     && \exp\left(-{1\over 2l^2}\sum_{I,i}(y_i^I)^2\right)
\end{eqnarray}
where the $Z_I$ are the center of mass coordinates:
\begin{equation}
  Z_I:=\sum_{k=1}^{N_I} z_k^I \,.
\end{equation}
{}From the quasiperiodicity of the $\vartheta_1$-functions
they derived the
following conditions on the center of mass functions
\begin{eqnarray}
   F^{c.m.}(Z^I+1) &=& F^{c.m}(Z^I) \\
   F^{c.m.}(Z^I+\tau)\!\!&=&\!\!\exp(-i\pi\tau K_{II}-2\pi i
   \sum_J K_{IJ}Z_J)F^{c.m}(Z^I)
\end{eqnarray}
and got $\abs{\det(K)}$ independent solutions for
$F^{c.m.}(Z_1,\ldots,Z_n)$
which gives a degeneracy of the groundstate of order $\abs{det(K)}$.
\pn
In order to study the influence on the boundary conditions to
the degenerate groundstate, we generalize the quasi
periodicity conditions and get for the center of mass functions:
\begin{eqnarray}
   F^{c.m.}(Z^I+1)\!\! &=&\!\! \exp(2\pi i\vpo)F^{c.m}(Z^I) \\
   F^{c.m.}(Z^I+\tau)\!\! &=&\!\! \exp(2\pi i \vpt)
   \exp(-i\pi\tau K_{II}-2\pi i \sum_J K_{IJ}Z_J)F^{c.m}(Z^I)\,.
\end{eqnarray}
The $\abs{\det(K)}$ solutions of $F^{c.m.}$ can be calculated in the
same way as before leading to the following expression:
\begin{eqnarray}
  F^{c.m.}_{\vec\alpha}(\vec Z) = \sum_{\vec m \in Z^n}\!\!\!
  &&\exp\left( i \pi\tau\vec m^t K \vec m + 2\pi i\tau \vec m
   (\vec\alpha+\vpo\vec e)-2\pi i \vec m \vpt\vec e\right)\\ \times
  &&\exp\left(2\pi i(\vec\alpha + K\vec m+\vpo\vec e\right) \vec Z)\,,
\end{eqnarray}
where $\vec e$ is the vector with entries all equal $1$ and
$\vec\alpha$ is a vector that characterizes the groundstate and
belongs to the coset ${\BZ^n / K \BZ^n}$ which has exactly
$\abs{\det(K)}$ elements.\pn
Let us make two remarks: Firstly, E.~Keski-Vakkuri and X.G.~Wen also
showed \cite{KW93} that the center-of-mass part $F^{c.m.}_{\vec\alpha}
(\vec Z)$ can be considered as the degenerate groundstate of a mean
field theory for the QHE, which is given by the following Chern-Simons
Lagrangian:
\begin{eqnarray}
   {\cal L}= {1\over 4\pi} \sum_{I,J} K_{IJ}\epsilon^{\mu\nu\lambda}
   a_{I\mu}\del_{\nu}a_{J\lambda}
\end{eqnarray}
This way of describing the QHE has been extensively studied
\cite{FK91,FZ91,LF90}.
\pn
Moreover,
the form of the $F^{c.m.}_{\vec\alpha}(\vec Z)$ does not
depend on the explicit structure of the wave function,
but only on the number
and degree of zeros of the wave function. This seems to be similar to the
Hofstadter problem which was recently reformulated by P.B.~Wiegmann and
A.V.~Zabrodin \cite{WZ93} who showed that the zeros of these wave
function fulfill some Bethe-Ansatz equations.  \pn
Finally, let us give an example.
\begin{eqnarray}
K=\left(
\begin{array}{cc}
 3 & 1	 \\
 1 & 3
\end{array}\right)
\Rightarrow
K^{-1}={1\over8}\left(
\begin{array}{rr}
 3	 & -1	 \\
 -1	 & 3
\end{array} \right)
\end{eqnarray}
Thus, $\det(K)=8$, $\nu={1\over 8}(3-1+3-1)={1\over 2}$ and the vectors
$\vec \alpha$ which label the eight-fold degenerate groundstate can be
written in the following way:
\begin{eqnarray}
\vec\alpha \in \left\{
\left(\begin{array}{c} 0\\ 0\end{array}\right),
\left(\begin{array}{c} 1\\ 1\end{array}\right),
\left(\begin{array}{c} 1\\ 2\end{array}\right),
\left(\begin{array}{c} 2\\ 1\end{array}\right),
\left(\begin{array}{c} 2\\ 2\end{array}\right),
\left(\begin{array}{c} 2\\ 3\end{array}\right),
\left(\begin{array}{c} 3\\ 2\end{array}\right),
\left(\begin{array}{c} 3\\ 3\end{array}\right)
\right\}
\end{eqnarray}
This may be represented graphically as follows.
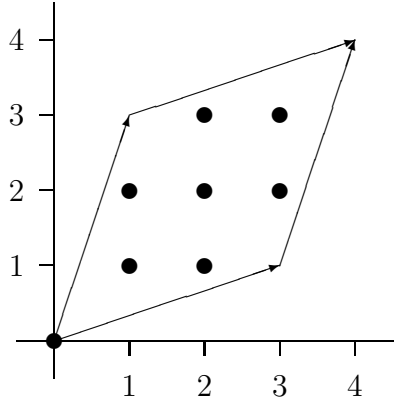
\begin{center}
   \unitlength=1mm
\special{em:linewidth 0.4pt}
\linethickness{0.4pt}
\begin{picture}(80.00,75.00)
\put(30.00,30.00){\line(1,0){50.00}}
\put(35.00,25.00){\line(0,1){50.00}}
\put(45.00,30.00){\line(0,-1){2.00}}
\put(55.00,30.00){\line(0,-1){2.00}}
\put(65.00,30.00){\line(0,-1){2.00}}
\put(75.00,30.00){\line(0,-1){2.00}}
\put(35.00,40.00){\line(-1,0){2.00}}
\put(35.00,60.00){\line(-1,0){2.00}}
\put(35.00,70.00){\line(-1,0){2.00}}
\put(35.00,30.00){\vector(3,1){30.00}}
\put(35.00,30.00){\vector(1,3){10.00}}
\put(45.00,60.00){\vector(3,1){30.00}}
\put(65.00,40.00){\vector(1,3){10.00}}
\put(45.00,40.00){\circle*{2.00}}
\put(55.00,40.00){\circle*{2.00}}
\put(55.00,60.00){\circle*{2.00}}
\put(65.00,60.00){\circle*{2.00}}
\put(35.00,30.00){\circle*{2.00}}
\put(45.00,24.00){\makebox(0,0)[cc]{1}}
\put(55.00,24.00){\makebox(0,0)[cc]{2}}
\put(65.00,24.00){\makebox(0,0)[cc]{3}}
\put(75.00,24.00){\makebox(0,0)[cc]{4}}
\put(30.00,40.00){\makebox(0,0)[cc]{1}}
\put(35.00,50.00){\line(-1,0){2.00}}
\put(45.00,50.00){\circle*{2.00}}
\put(55.00,50.00){\circle*{2.00}}
\put(65.00,50.00){\circle*{2.00}}
\put(30.00,50.00){\makebox(0,0)[cc]{2}}
\put(30.00,60.00){\makebox(0,0)[cc]{3}}
\put(30.00,70.00){\makebox(0,0)[cc]{4}}
\end{picture}
\end{center}
{}From now on let us assume that each subband contains the same
number of particles: $N_I=N_J=N,\quad I,J =1,\ldots,n$. From
equation (\ref{Kcond}) it follows, that $\vec e$ must be an
eigenvector corresponding to $K$.
Further, let us make the naturally restriction
to matrices $K$ which are invariant under arbitrary permutations
of all subbands. This symmetry implies that
$K_{IJ}=K_{\sigma(I)\sigma(J)}$ where $\sigma$ denotes an arbitrary
permutation of $\{1,\ldots,n\}$.
%%%%%%%%%%%%%%%%%%%%%%%%%%%%%%%%%%%%%%%%%%%%%%%%%%%%%%%%%%%%%%%%%%%%%%%%%
\mysection{Vector bundles over the Torus}
Next, we want to construct a
nontrivial vector bundle, which has the torus parameterized by $\vpo$ and
$\vpt$ as base manifold, from the generalized Laughlin wave functions with
periodic boundary conditions constructed in the last section.
Further, the filling factor can now be expressed as a topological quantity,
the first Chern number of the vector bundle divided by its rank. \pn
In the wave functions of the last section for a given matrix $K$
only the center-of-mass part of the wave function depends
on $\vpo$ and $\vpt$. Thus this part alone defines a vector bundle
by transition functions in the
following way:
\begin{eqnarray}
  \vpo\rightarrow\vpo+1 &:& F^{c.m.}_{\vec\alpha}(\vec Z)\rightarrow
  F^{c.m.}_{\vec\alpha+\vec e}(\vec Z) \\
  \vpt\rightarrow\vpt+1 &:& F^{c.m.}_{\vec\alpha}(\vec Z)\rightarrow
  F^{c.m.}_{\vec\alpha}(\vec Z) \\
  F^{c.m.}_{\vec\alpha+\vec{K_I}}(\vec Z)&=& \exp(\pi K_{II}+2\pi\alpha_I
  +2\pi(\vpo+i\vpt)) F^{c.m.}_{\vec\alpha}(\vec Z)\,,
\end{eqnarray}
where $\vec K_I$ is a column of the $K$.
The rank $r$ of this vector bundle is given by $\abs{\det(K)}$.
Refering to the example of the last section, the action of $\vpo$
can be represented graphically as follows:
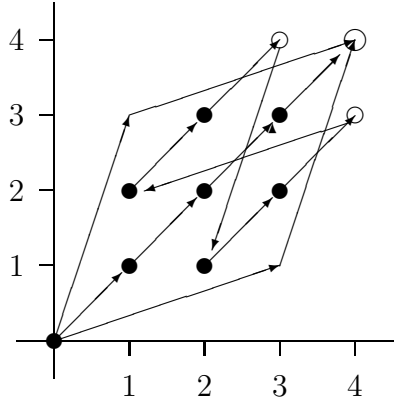
\begin{center}
  \unitlength=1.00mm
\special{em:linewidth 0.4pt}
\linethickness{0.4pt}
\begin{picture}(80.00,75.00)
\put(30.00,30.00){\line(1,0){50.00}}
\put(35.00,25.00){\line(0,1){50.00}}
\put(45.00,30.00){\line(0,-1){2.00}}
\put(55.00,30.00){\line(0,-1){2.00}}
\put(65.00,30.00){\line(0,-1){2.00}}
\put(75.00,30.00){\line(0,-1){2.00}}
\put(35.00,40.00){\line(-1,0){2.00}}
\put(35.00,60.00){\line(-1,0){2.00}}
\put(35.00,70.00){\line(-1,0){2.00}}
\put(35.00,30.00){\vector(3,1){30.00}}
\put(35.00,30.00){\vector(1,3){10.00}}
\put(45.00,60.00){\vector(3,1){30.00}}
\put(65.00,40.00){\vector(1,3){10.00}}
\put(45.00,40.00){\circle*{2.00}}
\put(55.00,40.00){\circle*{2.00}}
\put(55.00,60.00){\circle*{2.00}}
\put(65.00,60.00){\circle*{2.00}}
\put(35.00,30.00){\circle*{2.00}}
\put(45.00,24.00){\makebox(0,0)[cc]{1}}
\put(55.00,24.00){\makebox(0,0)[cc]{2}}
\put(65.00,24.00){\makebox(0,0)[cc]{3}}
\put(75.00,24.00){\makebox(0,0)[cc]{4}}
\put(30.00,40.00){\makebox(0,0)[cc]{1}}
\put(35.00,50.00){\line(-1,0){2.00}}
\put(45.00,50.00){\circle*{2.00}}
\put(55.00,50.00){\circle*{2.00}}
\put(65.00,50.00){\circle*{2.00}}
\put(30.00,50.00){\makebox(0,0)[cc]{2}}
\put(30.00,60.00){\makebox(0,0)[cc]{3}}
\put(30.00,70.00){\makebox(0,0)[cc]{4}}
\put(75.00,60.00){\circle{2.00}}
\put(75.00,70.00){\circle{2.83}}
\put(65.00,70.00){\circle{2.00}}
\put(55.00,60.00){\vector(1,1){10.00}}
\put(65.00,50.00){\vector(1,1){10.00}}
\put(35.00,30.00){\vector(1,1){9.00}}
\put(45.00,40.00){\vector(1,1){9.00}}
\put(55.00,50.00){\vector(1,1){9.00}}
\put(64.00,59.00){\vector(0,0){0.00}}
\put(65.00,60.00){\vector(1,1){8.00}}
\put(45.00,50.00){\vector(1,1){9.00}}
\put(55.00,40.00){\vector(1,1){9.00}}
\put(74.00,59.00){\vector(-3,-1){27.00}}
\put(65.00,69.00){\vector(-1,-3){9.00}}
\end{picture}

\end{center}
The Chern number or degree of the vector bundle can be
determined by counting the zeros of the corresponding determinant bundle
which is given by a product over a basis $\{ F^{c.m.}_{\vec\alpha_1},
\ldots, F^{c.m.}_{\vec\alpha_{\det(K)}} \}$. Then it is easy to
see that the first Chern number is given by
\begin{eqnarray}
   c_1=\abs{\det(K)}\abs{\sum_{I,J} (K^{-1})_{I,J}}\,.
\end{eqnarray}
Thus, the topological quantity
\begin{eqnarray}
   \mu = {c_1 \over r}
   \label{mu}
\end{eqnarray}
coincides with filling factor since $\nu$ is always positive if $\vec e$
is an eigenvector of $K$ (see equation (\ref{ff})).
\pn
In the following, we define an indecomposable vector bundle $E$ as a
vector bundle which can not be written as the direct sum of two other
bundles $E_1$ and $E_2$ over the same base manifold $M$: $E=E_1\oplus E_2$.
A vector bundle is called simple if $\dim_{\BBC}H^0(M,{\rm End}(E))=1$.
If $E$ is a simple vector bundle of rank $n$ over $M$ then $E$ is
indecomposable \cite{NS65}.
%and by the Riemann-Roch theorem for vector bundles
%$\dim_{\BBC}H^1(M,{\rm End}(E))=n^2(g-1)+1$ .
Studying now these vector bundles in more detail, it is not so difficult
to calculate that $\dim_{\BBC}H^0(M,{\rm End}(E))=\gcd(c_1,r) $. Thus,
the indecomposable vector bundles are determined by the condition
$\gcd(c_1,r)=1$ and given by the following matrices
\begin{eqnarray}
K=\left(\begin{array}{cccc}
 2p \pm 1	 & 2p		 & \ldots	 & 2p	 \\
 2p	  	 & \ddots		 & 		 & \vdots	 \\
 \vdots	 	 & 		 & 		 & 2p	 \\
 2p		 & \ldots	 & 2p		 & 2p\pm 1
\end{array}\right) \label{msm}
\end{eqnarray}
with filling fraction
\begin{eqnarray}
  \nu = {n \over 2pn \pm 1}\,.
\end{eqnarray}
Most remarkably, these are just half of the fractional values which
are found experimentally (\ref{ms}).
The electron-hole duality which transforms $\nu \rightarrow 1-\nu$
gives the other half of (\ref{ms}). Thus, we have found a topological
argument which selects the right fractional filling factors $\nu$
for the FQHE.
\pn
In the second section we have seen that the Hall conductivity is
calculated by averaging over the different boundary conditions. There is
also
another equivalent definition of the Chern number not counting the zeros
of the determinant bundle but using the curvature of the vector
bundle:
\begin{eqnarray}
   c_1 = {1 \over 2\pi i} \int d\varphi^2 {\rm tr}(F)\,.
\end{eqnarray}
The curvature is determined by the hermitian structure which is physical
naturally given by the scalar product which we already
used in the second section.
Comparing with equation (\ref{vector bundle}), $F$ is given by
\begin{eqnarray}
   F^{\vec{\alpha}\vec{\beta}}=
   \langle {\del \psi_{\vec\alpha}\over \del
   \vpo}\mid{\del \psi_{\vec\beta}\over \del \vpt}\rangle - \langle {\del
   \psi_{\vec\alpha}\over \del \vpt} \mid
   {\del \psi_{\vec\beta}\over \del \vpo}\rangle
\end{eqnarray}
where now $\psi_{\vec\alpha}$ are the normalized wave functions.
The explicit calculation of ${\rm tr} (F^{\vec{\alpha}\vec{\beta}})$
is very long and tedious.
One has to integrate over all electron coordinates $\{z_i^I\}$. But only
if one takes the whole electron wave function and not only the center of
mass term which already determines the vector bundle it is possible to
find an explicit expression for $F^{\vec\alpha\vec\beta}$. This is
technically very difficult since one has to handle with a lot of sums
coming from the $\vartheta_1$-functions in the wave function as can
be seen
in the Appendix A, but most surprisingly, it is possible to obtain an
explicit expression.
For $n=1$ and $K=(p)$ and more than two particles (i.e. $N>1$)
${\rm tr} (F^{\vec{\alpha}\vec{\beta}})$
can be expressed in the following way:
\begin{eqnarray}
  {\rm tr} (F^{\vec{\alpha}\vec{\beta}}) &=& {1\over Z}
  \bigsum{r_1,\ldots,r_{N-1}=0}{2pN-1}
  \alpha(r_1,\ldots,r_{N-1})\times \nonumber \\
  &&\sqrt{2\over pN} \sum_{k}\sum_{j=1}^{N}
  {1\over N}\exp\left(-{2\pi \over pN}(\vpo+k+r_j)^2 \right)\\
\end{eqnarray}
where the $\alpha(r_1,\ldots,r_{N-1})$ and $Z$ are given in the
appendix A. If $K$ is of the form (\ref{msm})
${\rm tr}(F^{\vec\alpha\vec\beta})$ can be expressed in a quite similar
way, $p$ has to be replaced by $\lambda$, the eigenvalue of $K$ by the
eigenvector $\vec e$.
${\rm tr}(F^{\vec\alpha\vec\beta})$ is independent of
$\vpt$ for more than two particles, but if it is plotted as a function
of $\vpo$ one sees that it oscillates or fluctuates around a mean value.
However, it is very remarkable that these fluctuations vanish
exponentially with the number of particles. That means
in the limit of an infinite particle number
${\rm tr}(F^{\vec\alpha\vec\beta})$
is independent of $\vpo$ and $\vpt$ and equals $2\pi i c_1$.
This is an important observation, since it explains the independence of
the QHE of boundary conditions and therefore the equality of exact
measurements for different sample geometries. This also justifies the
averaging of the Hall conductance in the second section in order to
express it as the first chern number of a vector bundle.
\pn
{}From a mathematical point of view,
the quantity $\nu$ is the most important quantity studying
stable vector bundles. Stability is an useful topological property
to restrict the moduli space of certain vector bundles. A vector bundle
$E$ is called stable if for every proper subbundle $F$ the relation
$\mu(F)<\mu(E)$ is true (or semi-stable if $\mu(F) \leq \mu(E)$).
Generally it is very difficult to check this property. But there is a
theorem of Donaldson that an indecomposable vector bundle over an
Riemannian surface is stable if and only if there exists a metric on the
vector bundle such that the trace of the curvature is constant \cite{D83}.
This metric is unique up to scale factors. In our cases on the torus
stability and indecomposability are equivalent.
\pn
It is very astonishing that just the Laughlin wave functions which
have been
found by numerical studies give the right hermitian structure for our
vector bundles, since only for them in the limit $N\rightarrow\infty$
${\rm tr}(F^{\vec{\alpha}\vec{\beta}})$
converges to a constant function.\pn
This shows a very nice interplay between experimental
physics, theoretical physics and pure mathematics.

\mysection{Acknowledgment}
I am very grateful to Werner Nahm for many illuminating discussions and
I would like to thank
Michael Flohr, Johannes Kellendonk,
Michael R\"osgen and Rudi Seiler for useful comments and also
Michael R\"osgen and Inga Daase for careful
reading of the manuscript. This work was supported by the Deutsche
Forschungsgemeinschaft.
%%%%%%%%%%%%%%%%%%%%%%%%%%%%%%%%%%%%%%%%%%%%%%%%%%%%%%%%%%%%%%%%%%%%%%%%
\pagebreak
\renewcommand{\theequation}{\mbox{\Alph{section}.\arabic{equation}}}
\begin{appendix}
\section{Appendix}
In the following, the multi summation indices are understand to have the
properties:
\begin{eqnarray}
   n_r^{ij} &=& -n_r^{ji}\,,\qquad 1\leq i,j \leq N\,,
                 \qquad r=1\dots p\\
   l_r^{ij} &=& -l_r^{ji}\,,\qquad 1\leq i,j \leq N\,,
                 \qquad r=1\dots p\\
   n^{ij}   &=& \sum_{r=1}^p n_r^{ij}\\
   l^{ij}   &=& \sum_{r=1}^p l_r^{ij}\\
   g^i      &=& \sum_{j\neq i}(n^{ij}+l^{ij}) \\
   {\rm erf}(x)&=& {2 \over\sqrt\pi } \int_0^x e^{-t^2}dt \,.
\end{eqnarray}
Then
\begin{eqnarray}
  <\tilde\psi_{\alpha}\!\mid\!\tilde\psi_{\beta}> &=& \delta_{\alpha\beta}
  \left({1 \over 8pN}\right)^{{N \over 2}}\exp\left({2\pi \over p}
  (\alpha+\vpo)^2\right)\times \nonumber \\
  \raisebox{5ex}{$\displaystyle \Bigsum
  {\begin{array}{c}
     m \in {\bf Z}  \\
     n^{ij}_1,\ldots,n^{ij}_p \in {\bf Z}+{1 \over 2}\\
     l^{ij}_1,\ldots,l^{ij}_p \in {\bf Z}+{1 \over 2}\\
     1\leq i<j\leq N \\
     \sum_{j=1}^{N}(n^{ij}-l^{ij})=0
  \end{array}}{}$}  \!\!\!\!\!&&
  \begin{array}{l}
  \\ \\
  \exp\bigg(-\pi\sum\limits_{i<j \atop r}
  \Big((n_r^{ij})^2 + (l_r^{ij})^2\Big)
  -{\pi \over 2Np}\Big(\sum\limits_i (g_i)^2\Big) \bigg)
  \nonumber \\
  \\
  \times
  \bigprod{i=1}{N}
  \bigg(
  {\rm erf}\Big({\pi \over 2pN}\big(2(\alpha+\vpo)+2p(m+N)+g_i\big)
            \Big)  \\
  \qquad\qquad
  - {\rm erf}\Big({\pi \over 2pN}\big(2(\alpha+\vpo)+2pm+g_i\big)
            \Big)
   \bigg)
  \end{array}
\end{eqnarray}
\begin{eqnarray}
  \Rightarrow\qquad
  <\tilde\psi_{\alpha}\!\mid\!\tilde\psi_{\beta}> = \delta_{\alpha\beta}
  {N \over (2pN)^{N \over 2}}\exp\left({2\pi \over p}
  (\alpha+\vpo)^2\right)
  \bigsum{r_1,\ldots,r_{N-1}=0}{2pN-1}
  \!\!\!\!\alpha(r_1,\ldots,r_{N-1})
\end{eqnarray}
\pagebreak
where
\begin{eqnarray}
   \alpha(r_1,\ldots,r_{N-1}) = \!\!\!\!\!
   \raisebox{-2ex}{$\displaystyle
   \Bigsum
   {\begin{array}{c}
   n^{ij}_1,\ldots,n^{ij}_p \\
   l^{ij}_1,\ldots,l^{ij}_p \\
   1\leq i<j\leq N \\
   n_1^{1i}=0\\
   \sum_{j=1}^{N}(n^{ij}-l^{ij})=0 \\
   g_i=r_i ({\rm mod\ }2pN)
   \end{array}}{} $}
  \!\!\!\!\!\!\!\!\!
  \exp\Big(-\pi\sum_{i<j \atop r} ((n_r^{ij})^2 + (l_r^{ij})^2)
  -{\pi \over 2Np}\big(\sum_i g_i^2\big)
  \Big)
\end{eqnarray}
and
\begin{eqnarray}
  Z:=
  \bigsum{r_1,\ldots,r_{N-1}=0}{2pN-1}\alpha(r_1,\ldots,r_{N-1})
\end{eqnarray}
With the definition of the connection
\begin{eqnarray}
  A_1^{\alpha\beta}&=&<\psi_{\alpha}\!\mid\!\del_{\vpo}\psi_{\beta}> \\
  A_2^{\alpha\beta}&=&<\psi_{\alpha}\!\mid\!\del_{\vpt}\psi_{\beta}> \\
  F^{\alpha\beta}  &=& \del_{\vpo}A_2^{\alpha\beta}-
                       \del_{\vpt}A_1^{\alpha\beta}
\end{eqnarray}
one gets
\begin{eqnarray}
  {1\over 2\pi i} {\rm tr}(\del_{\vpo}A_2^{\alpha\beta}) &=&
  {1\over Z}
  \bigsum{r_1,\ldots,r_{N-1}=0}{2pN-1}
  \alpha(r_1,\ldots,r_{N-1})\times \nonumber \\
  &&\sqrt{2\over pN} \sum_{k}\sum_{j=1}^{N}
  {1\over N}\exp\left(-{2\pi \over pN}(\vpo+k+{r_j\over 2})^2
  \right) \\
  {\rm with}\quad r_N &=& -r_1-\ldots-r_{N-1} \nonumber \\
  {1\over 2\pi i} {\rm tr}(\del_{\vpt}A_1^{\alpha\beta}) &=&
  0 \qquad {\rm if}\qquad N\geq 2
\end{eqnarray}
and
\begin{eqnarray}
  c_1&=&{1 \over 2\pi i}\int d^2\!\!\varphi\ {\rm tr}(F^{\alpha\beta}) =
  {1 \over 2N}{1 \over Z}
  \bigsum{r_1,\ldots,r_{N-1}=0}{2pN-1}
  \alpha(r_1,\ldots,r_{N-1})\times \nonumber \\
  &&\sum_{k}\sum_{j=1}^{N}
  \Big( {\rm erf}\big(\sqrt{2\pi \over pN}(k+1+{r_j\over 2})\big)
   - {\rm erf}\big(\sqrt{2\pi \over pN}(k+1+{r_j\over 2})\big) \Big)
  \nonumber \\
  &=& {1\over Z}
  \bigsum{r_1,\ldots,r_{N-1}=0}{2pN-1}
  \alpha(r_1,\ldots,r_{N-1}) = 1
\end{eqnarray}
\end{appendix}
%%%%%%%%%%%%%%%%%%%%%%%%%%%%%%%%%%%%%%%%%%%%%%%%%%%%%%%%%%%%%%%%%%%%%%%%%%%%
\def\refbf#1{\bf #1}
\def\AP#1{{ Ann.\ Phys.\ (N.Y.) {\refbf #1}}}
\def\APPB#1{{ Acta Phys.\ Polon.\ {\refbf B#1}}}
\def\ARNS#1{{ Ann.\ Rev.\ Nucl.\ Sci.\ {\refbf #1}}}
\def\ARNPS#1{{ Ann.\ Rev.\ Nucl.\ Part.\ Sci.\ {\refbf #1}}}
\def\DAN#1{{ Dokl.\ Akad.\ Nauk SSSR {\refbf #1}}} \let\DANS=\DAN
\def\CMP#1{{ Commun.\ Math.\ Phys.\ {\refbf #1}}}
\def\IMPA#1{{ Int.\ J.\ Mod.\ Phys.\ {\refbf A#1}}}
\def\IMPB#1{{ Int.\ J.\ Mod.\ Phys.\ {\refbf B#1}}}
\def\JDG#1{{J.\ Diff.\ Geom.\ {\refbf #1}}}
\def\JETP#1{{ Sov.\ Phys.\ JETP {\refbf #1}}}
\def\JETPL#1{{ JETP Lett.\ {\refbf #1}}}
\def\JMPA#1{{ Jour.\ of\ Mod.\ Phys.\ {\refbf A#1}}}
\def\JMP#1{{ J.\ Math.\ Phys.\ {\refbf #1}}}
\def\JSP#1{{ J.\ Stat.\ Phys.\ {\refbf #1}}}
\def\JPA#1{{ J.\ Phys.\ A:\ Math.\ Gen. {\refbf #1}}}
\def\LNC#1{{ Lett.\ Nuovo Cimento {\refbf #1}}}
\def\LMP#1{{ Lett.\ Math.\ Phys.\ {\refbf #1}}}
\def\MPL#1{{ Mod.\ Phys.\ Lett.\ {\refbf #1}}}
\def\NC#1{{ Nuovo Cimento {\refbf #1}}}
\def\NCA#1{{ Nuovo Cimento {\refbf #1A}}}
\def\NP#1{{ Nucl.\ Phys.\ {\refbf #1}}}
\def\NPB#1{{ Nucl.\ Phys.\ {\refbf B#1}}}
\def\PL#1{{ Phys.\ Lett.\ {\refbf #1}}}
\def\PLB#1{{ Phys.\ Lett.\ {\refbf #1B}}}
\def\PRep#1{{ Phys.\ Rep.\ {\refbf #1}}}
\def\PR#1{{ Phys.\ Rev.\ {\refbf #1}}}
\def\PRB#1{{ Phys.\ Rev.\ B\  {\refbf #1}}}
\def\PRD#1{{ Phys.\ Rev.\ D\  {\refbf #1}}}
\def\PRL#1{{ Phys.\ Rev.\ Lett.\ {\refbf #1}}}
\def\PRSA#1{{ Proc.\ Roy.\ Soc.\ London \ {\refbf A #1}}}
\def\PTP#1{{ Prog.\ Theor.\ Phys.\ {\refbf #1}}}
\def\PZETF#1{{ Pis'ma Zh.\ Eksp.\ Teor.\ Fiz.\ {\refbf #1}}}
\def\RMP#1{{ Rev.\ Mod.\ Phys.\ {\refbf #1}}}
\def\SJNP#1{{ Sov.\ J.\ Nucl.\ Phys.\ {\refbf #1}}}
\def\SJPN#1{{ Sov.\ J.\ Part.\ Nucl.\ {\refbf #1}}}
\def\SMD#1{{ Sov.\ Math.\ Dokl.\ {\refbf #1}}}
\def\SPD#1{{ Sov.\ Phys.\ Dokl.\ {\refbf #1}}}
\def\SPU#1{{ Sov.\ Phys.\ Usp.\ {\refbf #1}}}
\def\TMP#1{{ Theor.\ Math.\ Phys.\ {\refbf #1}}}
\def\UFN#1{{ Usp.\ Fiz.\ Nauk {\refbf #1}}}
\def\YF#1{{ Yad.\ Fiz.\ {\refbf #1}}}
\def\ZETF#1{{ Zh.\ Eksp.\ Teor.\ Fiz.\ {\refbf #1}}}
\def\ZPC#1{{ Z.\ Phys.\ C\ {\refbf #1}}}
\def\AM#1{{ Ann.\ Math.\ {\refbf #1}}}
\def\BAM#1{{ Bull.\ Am.\  Math.\ Soc.\ {\refbf #1}}}
\def\Phy#1{{ Physica\ {\refbf #1}}}
\def\IM#1{{ Invent.\ math.\ {\refbf #1}}}
\def\JPS#1{{ J.\ Phys.\ Soc.\ (Japan){\refbf #1}}}

\newpage
\begin{scriptsize}

\end{scriptsize}
%%%%%%%%%%%%%%%%%%%%%%%%%%%%%%%%%%%%%%%%%%%%%%%%%%%%%%%%%%%%%%%%%%%%%%%%%%%%%%

\begin{thebibliography}{99}
\bibitem{KDP80}  K.~Klitzing, G.~Dorda, M.~Pepper \PRL{45} (1980) 494
\bibitem{TSG82}  D.C.~Tsui, H.L.~Str\"omer, A.C. Gossard,
                 \PRL{48} (1982) 1559
\bibitem{J89a}   J.K.~Jain, \PRL{63} (1989) 199-202
\bibitem{DST93}  R.R.~Du, H.L.~Str\"omer, D.C.~Tsui, L.N.~Pfeiffer,
                 K.W.~West, \PRL{70} (1993) 2944
\bibitem{WRP93}  R.L.~Willet, R.R.,~Ruel, M.A.~Paalanen, K.W.~West,
                 L.N.~Pfeiffer, \PRB{47} (1993) 7344
\bibitem{HLR93}  B.I.~Halperin, P.A.~Lee, N.~Read, \PRB{47} 7312
\bibitem{FST94}   J.~Fr\"ohlich, U.~M.~Studer, E.~Thiran,
                 K.~U.~Leuven-preprint, May 1994, cond-mat/9406009
\bibitem{PG87}   R.E.~Prange, S.M.~Girvin, (1987)
                 {\it The Quantum Hall Effect}  (Berlin: Springer)
\bibitem{L81}    R.B.~Laughlin, \PRB{23} (1981) 5632-5623
\bibitem{TKN82}  D.J.~Thouless, M.~Kohmoto, M.P.~Nightingale,
                 M.~den~Nijs, \PRL{49}  (1982) 405-408
\bibitem{NTW85}  Q.~Niu, D.J.~Thouless, Y.S.~Wu, \PRB{31} (1985)
                 3372-3377
\bibitem{NT87}   Q.~Niu, D.J.~Thouless, \PRB{35,5} (1987) 2188-2197
\bibitem{ASS83}  J.E.~Avron, R.~Seiler, B.~Simon, \PRL{51} (1983) 51-53
\bibitem{AS85}   J.E.~Avron, R.~Seiler, \PRL{54} (1985) 259-262
\bibitem{K85}    M. ~Kohmoto, \AP{160} (1985) 343-354
\bibitem{N81}    S.P.~Novikov, \SMD{23} (1981), 298-303
\bibitem{L83}    R.B.~Laughlin, \PRL{50} (1983) 1395-1398
\bibitem{H83}    F.D.M.~Haldane, \PRL{51} (1983) 605-608
\bibitem{H84}    B.I.~Halperin, \PRL{52} (1984) 1583-1586
\bibitem{HR85}   F.D.M.~Haldane, E.H.~Rezayi, \PRB{31,4} (1985)
                 2529-2531
\bibitem{KW93}   E. Keski-Vakkuri, X.G.~Wen, Preprint CTP-2197,
                 hepth-9303155
\bibitem{D83}    S.K.~Donaldson, \JDG{18} (1983) 269-278
\bibitem{K57}    R.~Kubo, \JPS{12} (1957) 570
\bibitem{W92}    X.G.~Wen, \IMPB{6} (1992) 1711-1762
\bibitem{S91}    M.~Stone, \IMPB{5} (1991) 509
\bibitem{WZ92}   X.G.~Wen, A.~Zee, Preprint NST-ITP-92-10
\bibitem{J89b}   J.K.~Jain, \PRB{40} (1989) 8079-8082
\bibitem{WZ93}   P.B.~Wiegmann, A.V.~Zabrodin, Preprint LPTENS-93/34
\bibitem{FK91}   J.~Fr\"ohlich, T.~Kerler, \NPB{354} (1991) 369
\bibitem{FZ91}   J.~Fr\"ohlich, A.~Zee, \NPB{364} (1991) 517
\bibitem{LF90}   A.~Lopez, E.~Fradkin, \PRB{44} (1990) 5246
\bibitem{NS65}   M.S.~Narasimhan, C.S.~Seshadri, \AM{82} (1965) 540-567
%\bibitem{CTZ93}  A.~Cappelli, C.A.~Trugenberger, G.R.~Zemba,
%                 \NPB{396} (1993) 465-490, \PLB{306} (1993) 100-107
%\bibitem{FV94}   M.~Flohr, R.~Varnhagen, \JPA{27} (1994) 3999-4010
\end{thebibliography}
\end{document}